\documentclass[
  journal=largetwo,
  manuscript=article-type,
  year=2020,
  volume=37,
]{cup-journal}

\usepackage{amsmath}
\usepackage[nopatch]{microtype}
\usepackage{booktabs}
\usepackage{longtable}
\usepackage{subfigure}
\usepackage{graphicx}
\usepackage{pdfpages}
\usepackage{pgffor}
\usepackage{capt-of}

\graphicspath{{./figures/}}

\usepackage{comment}

\newcommand{\spirals}{S$\pi$RALS}
\newcommand{\kms}{km~s$^{-1}$}

\newcommand{\aap}{AAP}

\newcommand{\apjl}{ApJL}

\newcommand{\cols}{4}

\newcommand{\imgwidth}{0.23\linewidth}

\usepackage{courier}
\usepackage[percent]{overpic}
\usepackage[normalem]{ulem}

\title{Compact vs. Scattered: Suitability of Southern 6.7~GHz Methanol Masers for VLBI astrometry}

\author{L. J. Hyland}
\affiliation{School of Natural Sciences, University of Tasmania, Private Bag 37, Hobart, Tasmania 7001, Australia}
\email[L. J. Hyland]{Lucas.Hyland@utas.edu.edu}
\author{S. P. Ellingsen}
\affiliation{International Centre for Radio Astronomy Research, The University of Western Australia, 35 Stirling Highway, Crawley WA 6009, Australia} 
\author{M. J. Reid}
\affiliation{Center for Astrophysics $\vert$\ Harvard\ \&\ Smithsonian, Cambridge, MA\ 02138, USA}

 %% First letter not capped

\begin{document}

\begin{abstract}
 The 6.7~GHz methanol maser transition is exclusively associated with young, high-mass stars and represents a potential target for astrometric studies, including accurate determination of their distance through trigonometric parallax measurements. There are more than 1000 known 6.7~GHz methanol maser sources in the Milky Way; however, not all are suitable targets for astrometric measurements. We have used the Long Baseline Array to observe 187 southern 6.7~GHz methanol masers and identify 69 sources with one or more maser spots that are sufficiently compact and intense to be suitable targets for very long baseline interferometry astrometry with current instruments. Maser compactness appears to be a strong function of Galactic position, with masers that are likely in nearby spiral arms being more compact, while those associated with distant arms or the central Galactic region being less compact - a relationship we associate with scatter broadening. This has implications for astrophysical masers, especially distant ones employed for Galactic astrometry.
\end{abstract}

Accepted for publication in PASA.

\noindent 

\section{Introduction}
The scale and structure of our Galaxy, the Milky Way, is not well understood when compared to hundreds of other nearby galaxies. This is because the solar system's location in the plane of the Galaxy leads to significant obscuration at most wavelengths. Accurate distance determination is fundamental to many astrophysical investigations, and improving our knowledge of the distance to astronomical objects in the Milky Way is broadly beneficial. Trigonometric parallax (the apparent change of position of an object with respect to a distant background when viewed from different parts of the Earth's orbit about the sun), is the best method to measure distances to objects beyond the solar system. 

The Bar and Spiral Structure Legacy Survey \citep[BeSSeL;][]{Brunthaler2011,Reid2009f,Reid2014,Reid2019}, and the VLBI Exploration of Radio Astrometry \citep[VERA;][]{vera2020} project have used very long baseline interferometry (VLBI) arrays to measure trigonometric parallaxes towards more than 200 high-mass star-forming regions (HMSFRs) using water and methanol masers. As massive star formation regions are generally found within spiral arms, accurate distances allow us to map the spiral structure, constrain the rotation curve, determine stellar and gas dynamics, and measure the gravitational potential of the Milky Way.

Methanol masers have been grouped into two categories known as class I and class II \citep{Menten1991b}. Class I masing transitions are inverted when collisional processes dominate, while Class II masing transitions are inverted primarily through radiation \citep{Cragg1992}. The masers observed in the BeSSeL survey include Class II methanol transitions at 6.7 and 12.2~GHz as well as 22~GHz water masers. 

Our observations of southern sources are focused on  Class~II methanol masers at 6.7~GHz, which come from the $5_{1}\mbox{-}6_{0}\mbox{~A}^{+}$ transition \citep{Menten1991}. The 6.7~GHz methanol maser transition has been intensively studied for more than 30 years with more than 1000 sources detected in the Milky Way, primarily by the Methanol Multibeam catalogue \citep[MMB; ][]{Caswell2010,Green2010,Caswell2011,Green2012,Breen2015}. The MMB catalogue is the most complete survey of Southern Hemisphere ($l=186^\circ\rightarrow0^\circ\rightarrow60^\circ$) 6.7~GHz class II methanol masers to a depth of $3\sigma=0.51$~Jy \citep{Green2009,Green2017}. 

The 6.7~GHz methanol masers are the second brightest masing transition observed in interstellar space, only exceeded in intensity by the aforementioned 22~GHz water maser transition. Unlike the other most common centimetre wavelength maser species (OH and water), which can arise in a variety of astrophysical environments, the 6.7~GHz methanol masers are exclusively associated with young high-mass star formation regions \citep{Minier2003,Breen2013}. 

The intensity and compact nature of interstellar masers mean that they are generally excellent targets for VLBI observations. However, some 6.7~GHz class~II methanol masers can have the majority of their emission resolved at angular scales larger than 0.1 arcseconds \citep{Minier2002,Goedhart2005,Harvey-Smith2006,Bartkewicz2009}. The surveys that discover 6.7~GHz methanol masers are typically single-dish observations, and the intrinsic size of maser-emitting regions is orders of magnitude smaller than the instrument resolution. Follow-up interferometric observations are required to determine which sources have sufficiently strong emission on milliarcsecond scales to make them suitable targets for VLBI observations. 

The Southern Hemisphere Parallax Interferometric Radio Astrometry Legacy Survey (\spirals) is undertaking trigonometric parallax observations towards southern methanol masers to complement the BeSSeL Survey for the portions of the Galactic Plane that lie below a declination of approximately -30 degrees. At this declination, accurate astrometric observations become difficult or impossible for northern hemisphere interferometers. \spirals~ utilises the University of Tasmania (UTAS) VLBI Array consisting of the three 12m antennas from the AuScope geodetic array \citep{Lovell2013}, the Hobart~26m, the Ceduna~30m antenna \citep{McCulloch2005}, and the New Zealand Warkworth~30m telescope \citep{Woodburn2015}, for which the 6.7~GHz methanol transition is the most appropriate astrometric target.

In 2016, prior to the upgrade of the UTAS 12m telescopes allowing access to the 6.7~GHz line, we surveyed $\sim180$ southern 6.7~GHz methanol masers with the Australian Long Baseline Array (LBA) to determine which were suitable as targets for VLBI astrometry. In this paper, we report the results of that survey, including the statistics of 6.7~GHz methanol maser compactness that can be used for future southern VLBI experiments.

\section{Observations and Data Reduction}
We used the MMB to identify southern 6.7~GHz methanol masers in the Galactic longitude range $\ell = 188^{\circ} \rightarrow 360^{\circ}$ and latitude $|b| \le 2.5^{\circ}$ with a catalogued peak flux density $\ge10$~Jy. This gave 187 targets.

\begin{table}
	\centering
	\caption{VLBI baselines for the Australian LBA participating telescopes. {\bf Upper right:} Linear distances (km) between the antennas {\bf Lower left:} Approximate mean $uv$--distance (M$\lambda$) for our 6.7~GHz  observations. }
	\label{tab:baseline_table}
	\begin{tabular}{l| rrrrrr }
		\toprule
		\hline
		&{\bf At}&{\bf Cd}&{\bf Ho}		 &{\bf Mp}&{\bf Pa}&{\bf Wa} \\\hline
		{\bf At}&        &  1508  & 1396   		 &  114   &  322   & 2409     \\
		{\bf Cd}&   34   &        & 1702   		 & 1448   & 1361   & 3718     \\
		{\bf Ho}&   31   &    38  &        		 & 1286   & 1089   & 2415     \\
		{\bf Mp}&    2   &    32  &   29   		 &        &  207   & 2411     \\
		{\bf Pa}&    7   &    30  &   24   		 &    5   &        & 2425     \\
		{\bf Wa}&   54   &    83  &   54   		 &   54   &   54   &          \\ 
	   \bottomrule 
	\end{tabular}
 \end{table}

The observations were made using the LBA in two sessions on 2016 March 4 and 22 (project code V534). The array consisted of 6 antennas: the Australia Telescope Compact Array (AT), Murriyang~64m (PA) and Mopra~22m (MP) antennas operated by CSIRO, the Hobart~26m (HO) and Ceduna~30m (CD) antenna operated by UTAS, and the Warkworth~30m (WA) antenna operated by the Auckland University of Technology \citep{Woodburn2015}.  Baseline lengths are given in Table~\ref{tab:baseline_table}. The 6.7~GHz methanol maser targets were observed with scans of 150~seconds, augmented by 5~minute scans on fringe-finder quasars scheduled approximately every three hours to calibrate clock drift rates. Each maser target was observed three times over the two sessions. The data were recorded using the LBA Data Acquisition System (DAS), which recorded two IF bands, each 16~MHz dual circularly polarised, centered on 6308 and 6668~MHz sampled at a total data rate of 256~Mbits/s. Telescope baseband data were correlated with DiFX (Deller et al., 2007, 2011) using the LBA correlation facilities at the Pawsey supercomputer centre. The data for each observation were correlated in one pass with an integration time of 2~s and 8192 spectral channels. This gave a frequency resolution of 1.95~kHz (0.09 km s$^{-1}$ at 6.7~GHz).

The initial data reduction followed standard procedures for spectral-line VLBI using AIPS \citep{Greisen1990,Greisen2003}, including flagging bad data, amplitude and phase calibration, and velocity shifting to remove shifts from the Earth's orbit and rotation. The 6308~MHz band was only used to accurately measure the multi-band delay on the fringe finder quasars and remove the hydrogen maser drift rates at each telescope. Once calibrated, the visibility data (i.e., amplitude and phase for each polarisation, frequency, time and baseline) were further processed using python as described in the following section.

\subsection{Maser Visibility Fitting} \label{sec:analysis}
Interferometric imaging provides the most accurate information on the scale and distribution of the emission of the 6.7~GHz methanol masers, however, this requires good $uv$-coverage for each source. One alternative when there is insufficient data for good imaging is to fit a source model to the visibility amplitude, in order to estimate source size. This was commonly done in the early days of VLBI, when the number of antennas in arrays was small, or when brief observations had been made of large numbers of sources, as is the case here. \citet{Minier2002} were able to obtain reasonable fits for a number of 6.7 and 12.2~GHz methanol masers with a simple core-halo model, and here we follow a similar approach. The core is assumed to be angularly smaller than the halo. If we assume the flux density ($S$ in Jy) in each spectral channel is the combination of at least these two co-located Gaussian brightness distributions with FWHM $\theta_i$ (in rad) and some peak brightness $S_i$ (in Jy), then the resulting model for the total visibility as a function of baseline is:
\begin{equation*}
	S(uv) =~  S_He^{-\frac{2\pi^2}{8\ln 2}(\theta_H\,uv)^2} + S_ce^{-\frac{2\pi^2}{8\ln 2}(\theta_c\,uv)^2} 
\end{equation*} where $uv$ is the baseline length in units of the observing wavelengths (i.e., 4.45~cm), and subscripts of $H$/$c$ refer to the halo and core components, respectively. A maximum $uv$ baseline of $80-100$M$\lambda$ gives an angular resolution of $2.5-2$~mas on the sky, and as the baseline length approaches this limit, components with sizes much larger than this rapidly stop contributing significantly to the overall flux density. As we will see, any halo components rarely contribute outside of very short baselines ($uv<1-5$~M$\lambda$), so we will simplify the above equation to:
\begin{equation}
    \begin{split}
        \ln{S} \approx \ln{S_c} -3.56\theta_c^2 uv^2
    \end{split}
    \label{eq:model}
\end{equation} 
which is a straight line in $uv^2$ - $\log{S}$ space, the intercept giving the compact component flux density $S_c$ and the slope giving the compact core-size $\theta_c$. The most compact maser component(s) with flux density above the noise will dominate the observed trend, and the slope of the trend will reveal the size.

The position of most of the observed masers is not known to better than 0.1-1~arcsecond \citep{Green2008}, and in combination with the troposphere, the visibility phase can change sufficiently over 150~seconds to introduce decoherence, reducing the detected amplitude or making the maser completely undetectable in the amplitude spectrum (especially on the longer baselines). Therefore, we decided to fringe-fit each channel and recover the maser amplitude at the peak of the fringe-rate spectrum. This allows us to probe for maser emission down at the noise level of the interferometer, rather than averaging the visibility amplitudes incoherently \citep{TMS3}.

\begin{figure}
	\centering
	\includegraphics[width=\textwidth]{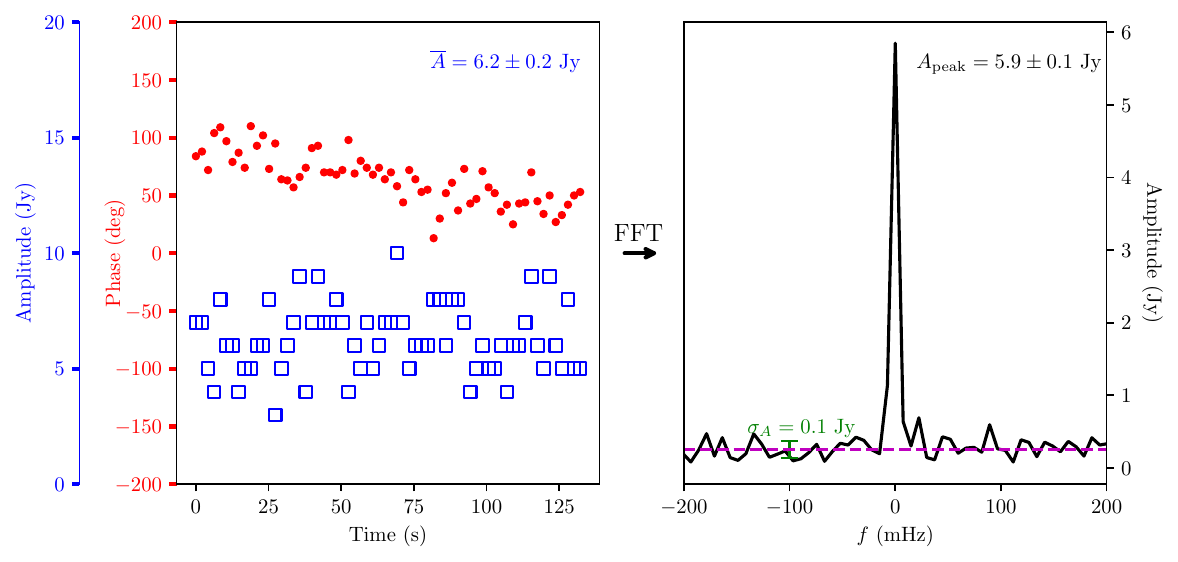}
    \includegraphics[width=\textwidth]{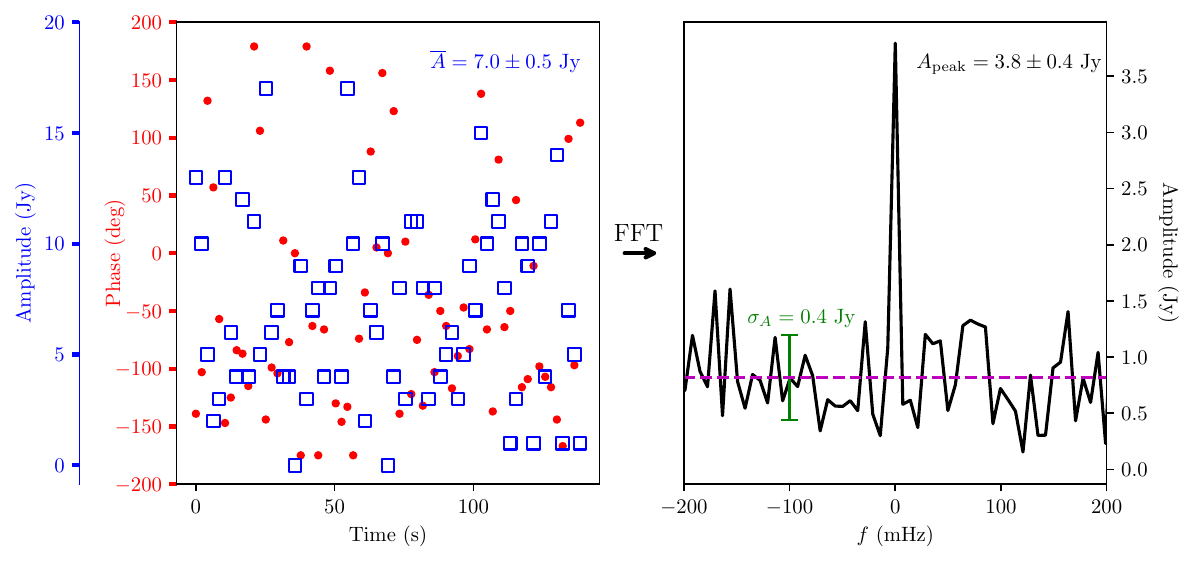}
    \caption{Example of the fringe-rate peak finding process for the maser 337.153-0.395. Visibility amplitude (blue squares) and phase (red dots) time series are on the left, while the fringe rate spectrum is on the right. \textbf{Top:} A clear detection in both the time and fringe rate domains on the CD-PA baseline. \textbf{Bottom:} A more tentative detection on the AT-WA baseline. The visibility time series does not have an obvious detection, but the fringe rate spectrum does. The green error bars in the fringe rate spectrum indicate the uncertainty in the peak measurement of the amplitude, while the purple dashed line represents the rice noise, which we are treating as the baseline noise threshold.}
	\label{fig:fft}
\end{figure}

For each maser, we selected a single 1.95~kHz channel that showed the most compact emission, determined either from the flux density on the longest baselines or by iterating the procedure described below. The visibility in that channel over time is:
\begin{equation*}
    V(t) = A(t)e^{i\phi(t)}=A(t)e^{i \left(ft + \phi_0\right)}
\end{equation*} where $f$ is the fringe-rate (Hz) and $\phi_0$ is an arbitrary phase offset (rad). Fourier transforming the visibility time series gives a fringe-rate spectrum for that channel:
\begin{equation*}
    \mathcal{F}\{V(t)\} = A(f) e^{i \phi(f)}
\end{equation*} which gives the flux density distributed across sampled fringe rates. The resolution of this spectrum is set by the scan length (150~s corresponds to 7~mHz), while the range of measurable fringe rates is limited by the time sampling (2~s corresponds to $\pm250$~mHz). Over a 150~s interval, position offsets and tropospheric fluctuations should contribute a near constant fringe-rate at 6.7~GHz. Thus, any true detection in the visibility time series should appear as a peak in the spectrum at the corresponding fringe rate. Figure~\ref{fig:fft} gives an example of this.

We searched for the strongest peak within the central $\pm100$ mHz of this spectrum (equivalent to a maximum offset of 4-arcsec from the correlated position of the maser for our longest baseline of 85~M$\lambda$), with the peak amplitude taken as the flux density in that channel on that baseline. We treat the $1\sigma$ scatter in the spectrum in the range $|f|>100$~mHz as the uncertainty in that flux density measurement, and the mean amplitude over the same range as the noise on the corresponding baseline for that scan (see Figure~\ref{fig:fft}). This process was repeated for all scans and baselines for the chosen maser channel. 

We adopted a 4$\sigma$ detection threshold; data with flux densities below this level were considered non-detections and omitted from further analysis. Using {\it scipy optimize} least squares, we fit the detected ${uv}^2$ vs. $\log S$ data to Equation~\ref{eq:model}, solving for the size and amplitude of any compact emission.

\subsection{Non-detections}
A total of 6 of the observed sources are considered not detected. Four of these; 294.337-1.706, 338.472+0.289, 338.850+0.409, and 338.902+0.394; had a catalogued flux density of less than 2~Jy and were not detected even in the Murriyang~64m autocorrelations. An incorrect position was used for the source 353.410-0.360, and so it was outside the primary beam of all telescopes. No observational fault can explain the missing 15~Jy maser 286.383-1.834, and while not previously known for this source, the most likely explanation is maser variability \citep{Caswell1995,Geodhart2004}. All other masers are detected in autocorrelations and on baselines $\le7$~M$\lambda$.

\subsection{Grading}
\begin{figure*}
    \centering
    \begin{overpic}[width=0.44\linewidth]{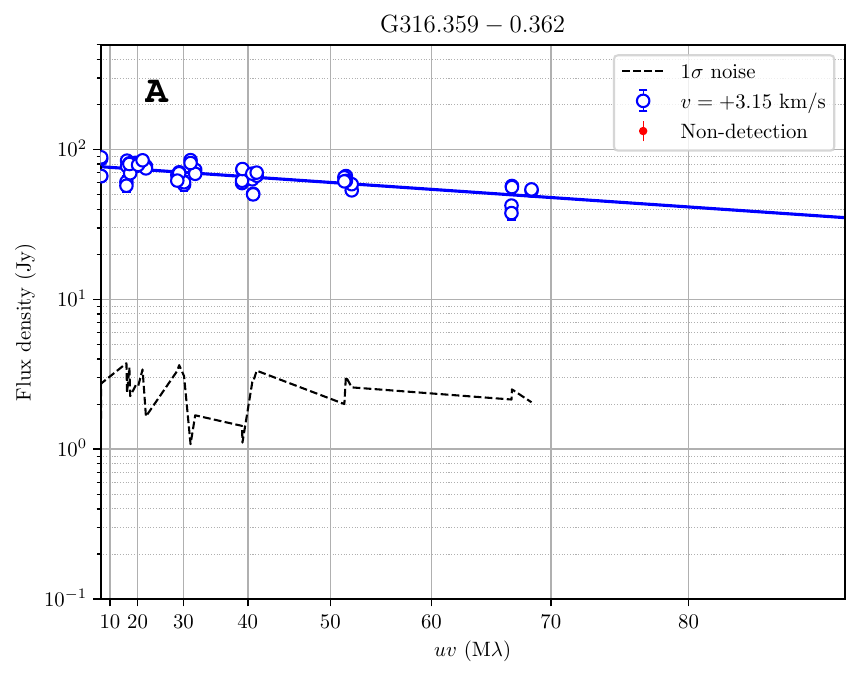}
        \put(17,65){\ttfamily \large \textbf{A}}
    \end{overpic}
    \begin{overpic}[width=0.44\linewidth]{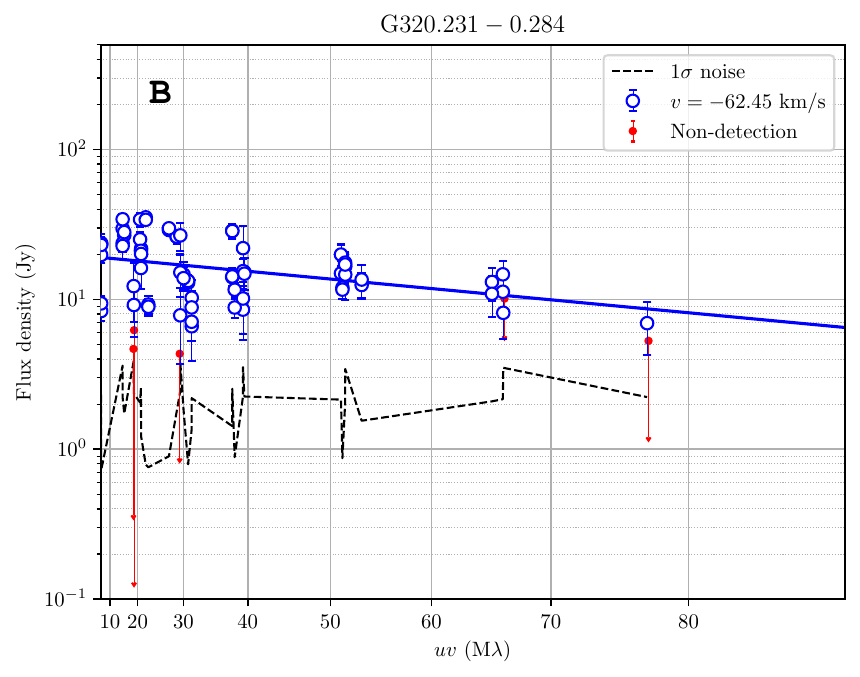}
        \put(17,65){\ttfamily \large \textbf{B}}
    \end{overpic}
    ~
    \begin{overpic}[width=0.44\linewidth]{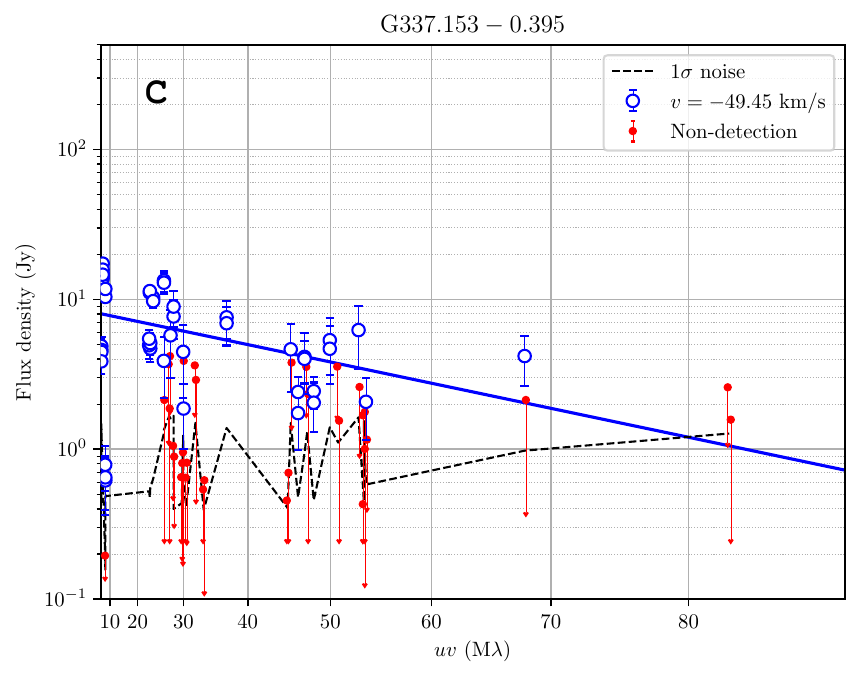}
        \put(17,65){\ttfamily \large \textbf{C}}
    \end{overpic}
    \begin{overpic}[width=0.44\linewidth]{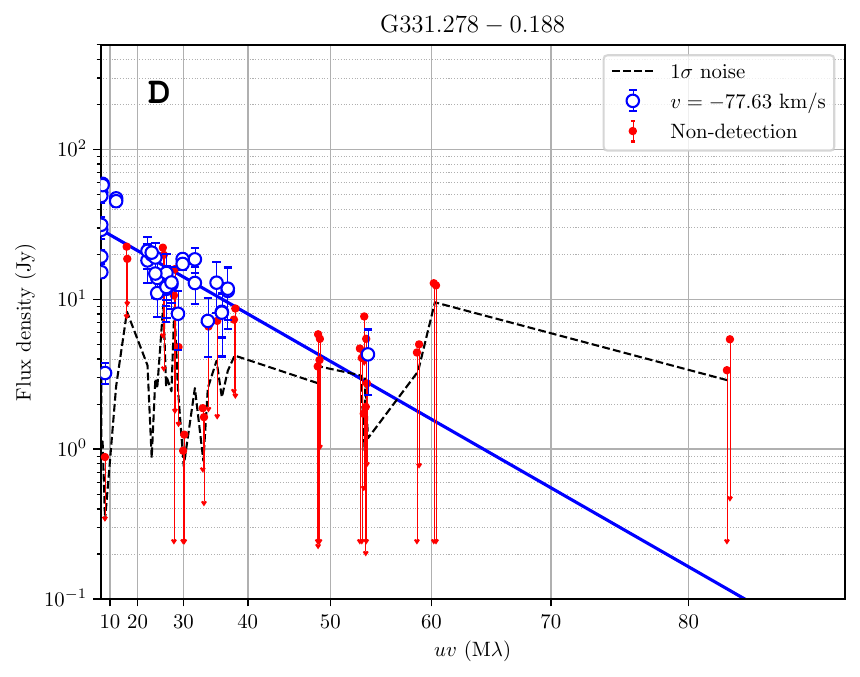}
        \put(17,65){\ttfamily \large \textbf{D}}
    \end{overpic}
    ~
    \begin{overpic}[width=0.44\linewidth]{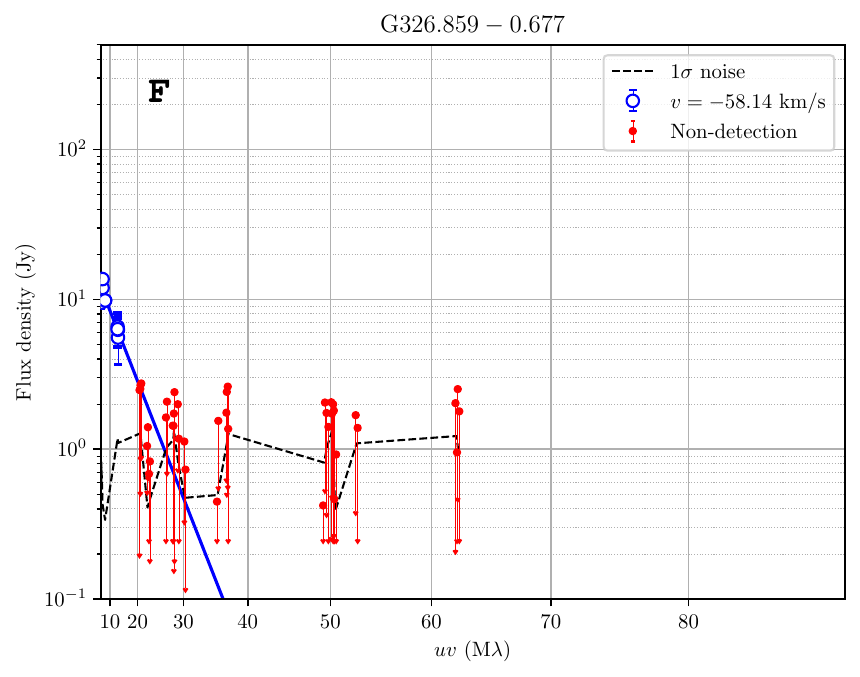}
        \put(17,65){\ttfamily \large \textbf{F}}
    \end{overpic}
    
    \caption{Examples of the 5 ``grades" of maser. Left to right, top to bottom: A, B, C, D and F. Blue points give the measured amplitude of the emission in the fringe-rate spectrum for the specified velocity channel, with the error bars giving the $3\sigma$ scatter in the fringe-rate spectrum, and the blue solid line is the fit to these data. Black dashed lines indicate the noise threshold in the fringe rate spectrum. Red points indicate detections below the acceptable threshold, which were not used in the fitting.}
    \label{fig:grades}
\end{figure*}

We visually inspected each maser and categorised them into 5 grades. Figure~\ref{fig:grades} shows an example of the 5 grades A to F in decreasing level of quality. Generally:
\begin{itemize}
    \item {\bf A}: Bright and compact. Clearly detected on all baselines.  
    \item {\bf B}: Compact, but $<10$Jy on longer baselines. 
    \item {\bf C}: Likely compact, but low flux density ($<10$~Jy). There are non-detections on longer baselines ($>50$M$\lambda$), and/or the detections on longer baselines are tentative, but the trend suggests that this is due to sensitivity.
    \item {\bf D}: Likely diffuse. Difficult to differentiate whether the emission is too diffuse or limited by sensitivity. Would require follow-up with a more sensitive array or imaging to confirm. 
    \item {\bf F}: Almost certainly diffuse. Clear detections on zero-spacing or short baselines, with a sharp fall off.
\end{itemize} The non-detected masers as discussed above have been categorised as N in Table~\ref{tab:detected_masers}.

\subsection{Fitting and Grading Results}
Of the 187 6.7 GHz methanol masers observed, we obtained useful data for 181 sources. Of these, the 51 F-grade masers are clearly not ideal for VLBI - a large proportion of their autocorrelated flux density is missing on the shortest ($2-7$~M$\lambda$ baselines), and they often fall below the noise threshold by 20~M$\lambda$. Any compact cores that may be present are below the noise floor (i.e., approximately $S_c<0.1-1$~Jy). 

There are a further 32 D-grade masers that are also likely not ideal. The flux density that is detected on short and intermediate baselines shows a trend that falls off much quicker than desired, likely indicative of large or weak cores ($\ge3$~mas). There are no convincing detections on long baselines. The low flux density of the emission makes it unclear whether the observed drop-off is due to resolution or sensitivity. 

The 29 C-grade masers are very similar to the previous D-grade, however, the observed trend is much shallower due to detection on intermediate-longer baselines ($60-70$~M$\lambda$) and indicates a smaller core size ($0.1-2$~mas). As with the D-grade masers, the low flux density makes it difficult to conclusively determine that they are compact - but there is evidence to suggest that they are. 

The final two grades are the `good' A and B masers. The 32 A-grade masers are clearly detected well-above the noise threshold on every baseline, with a shallow trend indicating a small core size. The remaining 37 B-grade masers often have non-detections on some baselines, but are either clearly detected on a long baseline ($>70$~M$\lambda$) with a shallow trend, or detected on almost all baselines with an upper cutoff of 10~Jy. 

Table~\ref{tab:compact_gradenumber} shows the number and fraction of masers within each grade, and the $uv^2$ vs. $\log S$ plots for all masers are included in the Appendix.

\begin{table}[h]
    \centering
    \begin{tabular}{c|cccccc}
        \hline
        {\bf Grade} & {\bf A} & {\bf B} & {\bf C} & {\bf D} & {\bf F} & {\bf N} \\\hline
        {\bf Number} & 32 & 37 & 29 & 32 & 51 & 6 \\
        {\bf Fraction} & 0.17 & 0.20 & 0.16 & 0.17 & 0.27 & 0.03 \\\hline
    \end{tabular}
    \caption{Number and fraction of each maser grade.}
    \label{tab:compact_gradenumber}
\end{table}

\subsection{Galactic distribution of compact masers}

\begin{figure*}
	\centering
	\includegraphics[width=0.98\textwidth]{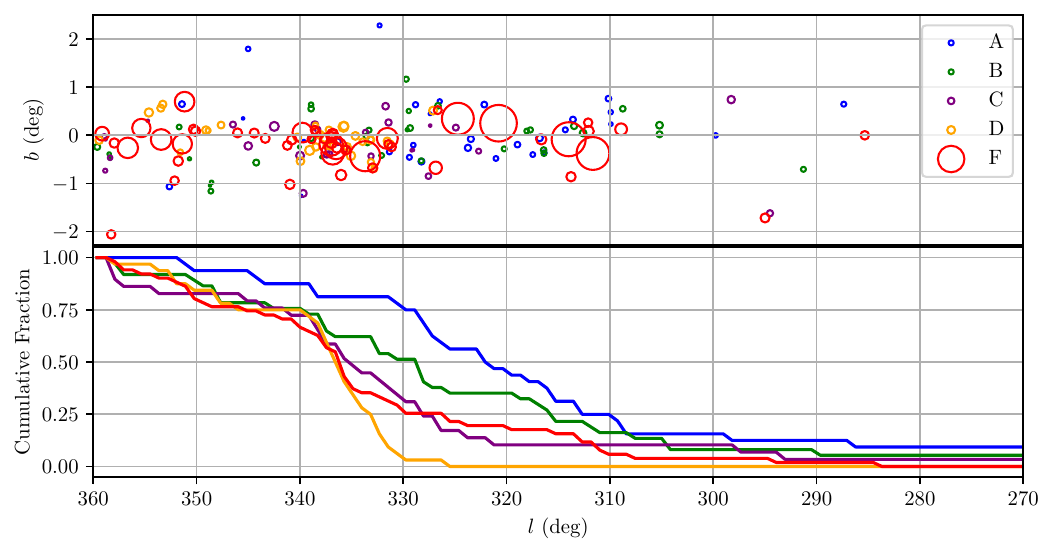}
	\caption{\textbf{Top:} Distribution of maser grade and size vs. Galactic longitude (between 270 and 360~deg) and Galactic latitude. \textbf{Bottom:} Cumulative fraction of each maser grade vs. increasing Galactic longitude.}
	\label{fig:sizevslb}
\end{figure*}

\begin{figure*}
	\centering
	\includegraphics[width=0.98\textwidth]{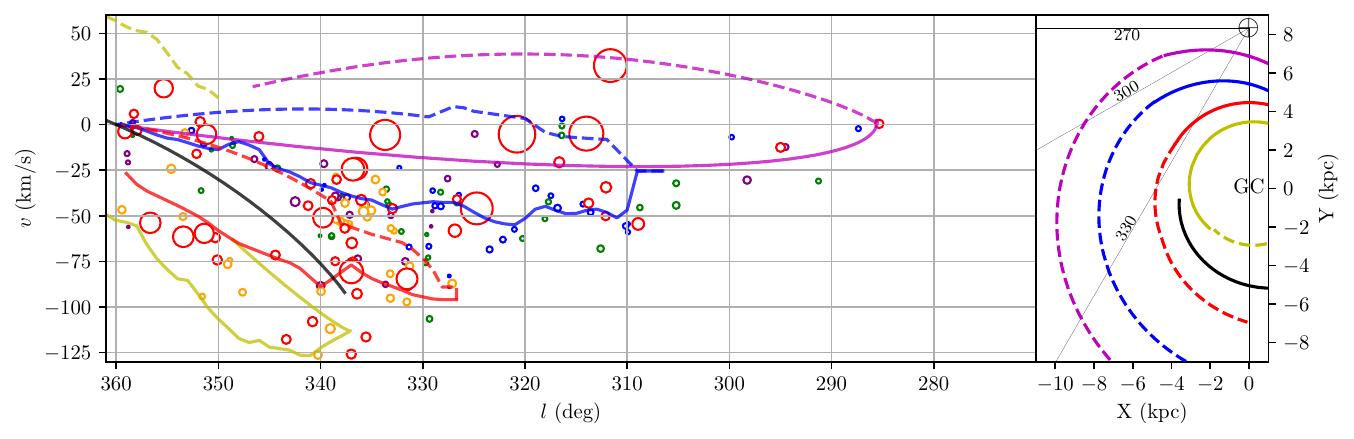}
	\caption{\textbf{Left:} Distribution of maser grade and size on a longitude vs. velocity diagram with spiral arms from \cite{Reid2019}. Maser grades are coloured as before. \textbf{Right:} The same arms on a plan view of the Milky Way for reference. Arms are coloured as: Carina (purple), Centaurus (blue), Norma (red), "3-kpc" (yellow), and Perseus (black). Dashed lines indicate arms past the tangent point. Galactic longitudes between $270\rightarrow360$ in increments of 30~deg are shown with thin black lines, the Sun is shown at (0,8.15~kpc) with the sun-symbol, and the Galactic centre is shown as "GC".} 
	\label{fig:sizevslv}
\end{figure*}

\begin{figure}
	\centering
	\includegraphics[width=0.98\textwidth]{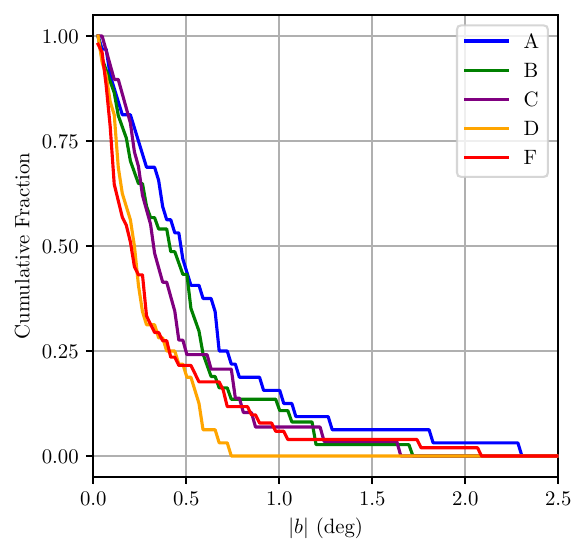}
	\caption{Cumulative fraction of each maser grade against decreasing absolute Galactic latitude $|b|$. More compact masers are preferentially found at larger Galactic latitudes than less compact masers.}
	\label{fig:gradevsb}
\end{figure}

Figure~\ref{fig:sizevslb} shows the distribution of measured maser size, grade and cumulative fraction of grade against Galactic position, Figure~\ref{fig:sizevslv} shows the same distribution against Galactic latitude and velocity with expected spiral arm positions \citep{Reid2019}, and Figure~\ref{fig:gradevsb} shows the cumulative fraction of each grade against total Galactic latitude $|b|$.

Between Galactic longitudes 305 and 360~deg, A and B-grade masers appear evenly distributed, with 50\% of them between 310-330~deg. Conversely, the distribution of D and F-grade masers peaks much later, showing jumps between 330-340~deg, and 350-360~deg. The C-grade masers appear somewhat different, with a steady increase after 320~deg and more closely following the A and B trends. It is difficult to draw conclusions about the outer Galactic regions $l<305$~deg due to the lack of masers.

In Figure~\ref{fig:gradevsb}, as we trend closer to the Galactic mid-plane ($b=2.5\rightarrow0$~deg), the order that masers reach 50\% total fraction goes A, B, C, D, and F. The point at which each grade of maser reaches 50\% of the total fraction is: A 0.48~deg, B 0.42~deg, C 0.33~deg, D 0.24~deg and F 0.22~deg. There is also a noticeable bump in the number of D-grade masers within 0.5~deg.

Comparing these trends to Figure~\ref{fig:sizevslv}, we can see 330-340~deg coincides with almost every spiral arm and the predicted edge of the far-bar. Almost all of the A and B sources appear to be associated with the near Centaurus or Carina arms. As the A-grade masers are generally quite bright, this is consistent with expectations; however, the weaker B-grade masers would be expected to be distributed randomly throughout the Galaxy if the size is mostly intrinsic. Almost all sources associated with the central 3-kpc region (yellow) are either D or F; the exceptions are the C-grade masers 358.809-0.085 or 339u.986-0.425. The single source from our sample that is clearly associated with the far Centaurus arm 311.643-0.380, is an F-grade maser. It is difficult to draw any conclusions about the 325-340~deg, $-25<v<-50$~\kms~ region in Figure~\ref{fig:sizevslv} due to the alignment of the far Norma and near Carina arms.

\section{Discussion}
\subsection{Maser Variability and Flaring}
The vast majority of the masers we observed had a measured flux density within 10-30\% of the MMB catalogued flux density, similar to the expected amount due to variability \citep{Caswell1995} and within our amplitude calibration uncertainty. The three exceptions were the non-detected maser 286.383-1.834 as discussed above, 192.600-0.048 (aka S255) and 323.459-0.079 (G323.4). 

S255 was undergoing a massive flare at the time of our observations, as reported in \citet{Szymczak2018b}. Our observations were undertaken on MJD 57451 and 57469, very close to the peak of the flare in the 5.2~km/s at MJD 57477. Pre-flare, the peak flux density of the maser was $\sim80$~Jy \citep{Caswell1995a} and during the flare the peak reached 1500~Jy. The maser underwent significant spectral changes, then returned to approximately pre-flare flux densities within a few years. Our results show the flaring regions were all very compact and the maser was categorised as A-grade, and this compactness is reflected by successful VLBI imaging \citep{Moscadelli2017}.

The catalogued flux density for G323.4 is about 20~Jy \citep{Caswell1995a,Green2012}, however, we detected the emission at approximately 500~Jy. G323.4 is now known to be a variable maser that is thought to undergo periodic flaring \citep{Proven-Adzri2019}, and the original large flare, which likely caused the as-so-far permanent spectral changes, is now thought to have occurred in 2013 \citep{Wolf2024}. The maser has since been reported in a much lower flux density state of 70~Jy. 

Since our data for these masers is exclusively from the flaring period, our flux density vs. baseline results would not reflect the current states. Our results for other masers that have since undergone known flaring, such as 358.931-0.030 \citep{Sugiyama2019}, should be treated with care.% However, if the dominant effect of 6.7~GHz maser compactness is due to the interstellar medium (see below), then any maser we have measured to be compact should remain so in principle. 

\subsection{Interstellar scattering}
Multi-path diffraction through the Interstellar Medium (ISM) causes scintillation and angular broadening, aka interstellar scattering \citep{Cordes1991b,Fey1991,Pushkarev2015}, which scales with frequency as $\theta\propto\nu^{-2}$. We postulate that while both cores and halos are intrinsic structures in masers, the small maser cores that are further away are more likely to be affected by interstellar scattering. Conversely, from a geometric perspective, maser cores that are further away should be more compact.

Figure~\ref{fig:sizevslb} shows that 75\% of the undoubtedly compact (A and B) masers are found outside of $l=330$~deg, while 75\% of the F and nearly 100\% of the D-grade masers are found within 330-360~deg. This is further highlighted in Figure~\ref{fig:sizevslv}, where almost all the compact masers are associated with the expected position of the near Centaurus arm at about $d=3$~kpc. 

As the scale height of masers in the Milky Way is 19~pc \citep{Reid2019}, masers that are further away will have a smaller Galactic latitude away from regions with Galactic warping. It follows that masers with a larger $b$ are more likely to be nearby. Masers that are seen through the Galactic plane would likely be more affected by interstellar scattering than masers closer to or not seen through the Galactic plane. Figure~\ref{fig:gradevsb} shows a trend that masers that are less compact are closer to the Galactic plane than masers that are more compact, implying that distant masers are less compact than nearby ones. Only a handful of sources (e.g., with $|b| > 1$~deg) are likely in warped regions, and these should not significantly affect our analysis.

These data show a clear trend that masers that are further away, by proxy of Galactic position and/or spiral arm association, are less compact than those closer. 

The ISM is known to be inhomogeneous and irregular \citep{Rickett1990}, which we propose to be clumpy with holes. Therefore, we present 3 cases for the size of maser cores:
\begin{enumerate}
    \item No or weak compact cores (halo-dominated): The maser is intrinsically large. Regardless of distance or Galactic position (especially if they are larger than the scattering size), the maser will appear large and is unsuitable for VLBI. This likely holds for the 12.2~GHz methanol transition, as they share similar mechanisms and structural characteristics \citep{Minier2002}. It is unclear whether 22~GHz water maser astrometric observations would be successful. 
    \item The maser has a compact core, and appears so. The maser is either close or viewed through holes in the ISM, and thereby remains compact. In either case, it is a good target for astrometric VLBI. The 12.2~GHz methanol and 22~GHz water masers should also be compact.
    \item The maser has a compact physical structure, but due to interstellar scattering, it has been angularly broadened. This is more likely to happen if the maser is far away. Observations of the 12.2~GHz methanol or 22~GHz water transitions would lessen the effect of interstellar scattering (as it scales as $\nu^{-2}$) and may allow VLBI astrometric observations. 
\end{enumerate}

There have been previous VLBI observations investigating these phenomena. \citet{Pushkarev2015} measured VLBI quasar size as a function of Galactic latitude and frequency. They found that there was a significant difference in the modelled angular size of AGN at lower latitudes ($|b|<10^\circ$) and those well above the Galactic plane ($|b|>10^\circ$). Their 2 and 8~GHz data were collected simultaneously, allowing for the frequency dependence to be explored. For sources at lower Galactic latitudes, 33\% had a frequency-dependent core size trending as $\nu^{-2}$, suggesting scatter broadening.

\citet{Minier2002} investigated whether the presence of existence of halos around maser cores can be explained by scattering broadening. They argue that while some degree of scattering is expected, it cannot give rise to the 5-50~mas halo structures, and that the ratio of halo size between 6.7 and 12.2~GHz did not behave as $\nu^{-2}$. Our hypothesis is consistent with their results on halos, and they do not compare the sizes of the 6.7~GHz and 12.2~GHz cores.

\citet{Menten1992} directly compared the cores of 6.7~GHz methanol, 12.2~GHz methanol, and 1.665~GHz OH at the same velocities in W3(OH). Their conclusion was that since a $\nu^{-2}$ variation was not seen, the observed spot size was intrinsic. W3(OH) is now known to be a close maser source \citep[$d=2.04$~kpc;][]{Hachisuka2006}. It is also in the outer Galaxy ($l=134$~deg) where the electron density is lower \citep{Cordes1991b}. We suggest that, due to these two factors and as the authors claim, W3(OH) is indeed not affected by scattering (i.e., case 2 above). Despite this, the method of comparing core sizes at different frequencies has the potential to further investigate our claim of systematic scatter broadening, and discover which distant HMSFRs do host compact cores suitable for VLBI astrometry (case 3). 

To discover these case 3 masers, follow-up observations are required. The most suitable transitions for current Southern Hemisphere facilities are the 12.2~GHz methanol masers and the 22~GHz water masers. Both transitions have been observed with great success in the Northern Hemisphere, particularly the 22~GHz water line, with parallax distances further than 10~kpc \citep{Reid2019,vera2020}. Conversely, the 12.2~GHz methanol line may be the most suitable for southern hemisphere VLBI observations, as fewer radio telescopes have access to the 22~GHz band. 

\section{Conclusion}
We have made very long baseline observations of 187 southern 6.7~GHz methanol maser sources, the largest systematic survey of these objects at southern latitudes. Of these, 69 are sufficiently compact and intense that they can be used for astrometric observations with current generation instruments, with a further 29 likely useful for observations with VLBI arrays that include larger apertures like MeerKAT phase 3 or SKA-MID AA*. This implies that $\ge54\%$ of 6.7~GHz methanol masers with a peak flux density above 10~Jy are suitable for VLBI observations. Additionally, follow-up observations of C- and D-grade masers with these next-generation facilities, including more sources down to the 1~Jy threshold, may reveal more VLBI candidates. This appears particularly important for the Galactic centre and for the distant spiral arms, where there are very few sources that we have determined have compact cores. The size of 6.7~GHz masers appears to be a combination of intrinsic factors and interstellar scattering, with the latter becoming dominant for sources within the Galactic centre and (based on spiral arm assignment) past 5~kpc. For these HMSFR with broadened 6.7~GHz maser emission, VLBI observations at higher frequencies (12.2 or 22~GHz) may be an alternative way to measure the distance.

\begin{acknowledgement}
The LBA is part of the Australia Telescope National Facility, which is funded by the Australian Government for operation as a National Facility managed by CSIRO and the University of Tasmania. This research has made use of NASA's Astrophysics Data System Abstract Service. This research made use of \url{MaserDB.net}, an online database of astrophysical masers \citep{maserdb2019}. This research made use of Astropy, a community-developed core Python package for Astronomy \citep{astropy:2013,astropy:2018}. We acknowledge the Gamilaroi (MP), Gomeroi (AT), Paredarerme (HO), Wiradjuri (PA) and Wiriangu (CD) peoples as the traditional caretakers of the land situating the Australian radio telescopes used in this work.
\end{acknowledgement}

\paragraph{Funding Statement}
This research was supported by the Australian Research Council (ARC) Discovery grants No. DP180101061 and DP230100727. 

\paragraph{Competing Interests}
None.

\paragraph{Data Availability Statement}
The correlator FITS files and associated metadata for the Long Baseline Array data underlying this article are available via the Australia Telescope Online Archive under experiment code V534.

\printendnotes

\printbibliography

\appendix
{\onecolumn
\section{Maser Flux Densities, Sizes and Grades}

\begin{longtable}{lrrrr|rrrrrrr|c}
      \caption{Flux densities, sizes and grades for masers observed in this work. \textbf{Columns: (1)} Source name in Galactic coordinates; \textbf{(2)} Right Ascension (J2000); \textbf{(3)} Declination (J2000); \textbf{(4)}; maser spot LSR (km/s); \textbf{(5)} angular size of core (mas);\textbf{(6)} flux density of core (Jy); \textbf{(7-12)} flux density of maser component (Jy) in the $uv$ range 0-10, 10-35, 35-60, 60-70, 70-80, 80-90~M$\lambda$ (if applicable) \textbf{(13)} maser grade.}
      \label{tab:detected_masers} \\
      \hline \\[-2ex]
      \textbf{Name}&\textbf{RA (J2000)}&\textbf{DEC (J2000)}&\multicolumn{1}{c}{$\boldsymbol{V}$} & \multicolumn{1}{c}{$\boldsymbol{\theta_c}$}&\multicolumn{1}{c}{$\boldsymbol{S_c}$}&
      $\boldsymbol{S_{>0-10}}$&$\boldsymbol{S_{10-35}}$&$\boldsymbol{S_{35-60}}$&
      $\boldsymbol{S_{60-70}}$&$\boldsymbol{S_{70-80}}$&$\boldsymbol{S_{80-90}}$&\textbf{Grade}\\ 
      &\textbf{hh:mm:ss}&\textbf{dd:mm:ss}&\textbf{(\kms)}&\textbf{(mas)}&\multicolumn{1}{c}{\textbf{(Jy)}}&\multicolumn{1}{c}{\textbf{(Jy)}}&\multicolumn{1}{c}{\textbf{(Jy)}}&\multicolumn{1}{c}{\textbf{(Jy)}}&\multicolumn{1}{c}{\textbf{(Jy)}}&\multicolumn{1}{c}{\textbf{(Jy)}} &\multicolumn{1}{c}{\textbf{(Jy)}} & \\\midrule
     \endfirsthead   
     \multicolumn{4}{c}{{\bf \tablename}  {\bf \thetable{}} -- Continued...} \\[0.5ex]
     \hline \\[-2ex]
    \textbf{Name}&\textbf{RA (J2000)}&\textbf{DEC (J2000)}&\multicolumn{1}{c}{$\boldsymbol{V}$} & \multicolumn{1}{c}{$\boldsymbol{\theta_c}$}&\multicolumn{1}{c}{$\boldsymbol{S_c}$}&
      $\boldsymbol{S_{>0-10}}$&$\boldsymbol{S_{10-35}}$&$\boldsymbol{S_{35-60}}$&
      $\boldsymbol{S_{60-70}}$&$\boldsymbol{S_{70-80}}$&$\boldsymbol{S_{80-90}}$&\textbf{Grade}\\ 
      &\textbf{hh:mm:ss}&\textbf{dd:mm:ss}&\textbf{(\kms)}&\textbf{(mas)}&\multicolumn{1}{c}{\textbf{(Jy)}}&\multicolumn{1}{c}{\textbf{(Jy)}}&\multicolumn{1}{c}{\textbf{(Jy)}}&\multicolumn{1}{c}{\textbf{(Jy)}}&\multicolumn{1}{c}{\textbf{(Jy)}}&\multicolumn{1}{c}{\textbf{(Jy)}} &\multicolumn{1}{c}{\textbf{(Jy)}} & \\\midrule\\[-1.8ex]
     \endhead
     \\[-0.1ex] \hline
     \multicolumn{7}{l}{{Continued on Next Page\ldots}} \\
     \endfoot
     \\ \bottomrule
     \endlastfoot
188.946+0.886& 06:08:53.32&  21:38:29.1&   +10.61&    2.1&  157.4&  425.7&  149.6&   47.9&       &       &   33.4&  A \\
189.030+0.783& 06:08:40.65&  21:31:07.0&    +8.72&    3.1&   12.6&   14.6&    7.3&    2.9&       &       &       &  C \\
192.600-0.048& 06:12:53.99&  17:59:23.7&    +5.08&    1.4&   69.5&   98.2&   58.3&   43.4&       &       &   28.8&  A \\
196.454-1.677& 06:14:37.03&  13:49:36.6&   +14.47&    1.7&    6.1&    7.9&    5.5&    3.5&       &       &    2.0&  B \\
232.620+0.996& 07:32:09.79& -16:58:12.4&   +22.56&    2.2&  126.1&  144.1&   86.9&   49.6&   15.0&   12.6&       &  A \\
263.250+0.514& 08:48:47.84& -42:54:28.3&   +12.09&    2.0&   31.6&   38.9&   22.9&   19.2&   10.9&       &       &  B \\
285.337-0.002& 10:32:09.62& -58:02:04.6&    +0.43&    3.4&    7.9&    7.7&    4.5&       &       &       &       &  F \\
286.383-1.834& 10:31:55.12& -60:08:38.6&   +14.29&    0.0&   13.8&   12.4&   14.5&   14.3&       &       &       &  N \\
287.371+0.644& 10:48:04.44& -58:27:01.0&    -2.21&    1.2&   41.9&   79.2&   28.1&   44.7&   25.0&       &   15.4&  A \\
291.274-0.709& 11:11:53.35& -61:18:23.7&   -30.93&    1.2&   24.8&   46.9&   18.4&   23.7&   20.6&       &       &  B \\
294.337-1.706& 11:33:49.91& -63:16:32.5&    -6.52&    1.6&    0.6&    0.7&    0.5&       &       &       &       &  N \\
294.511-1.621& 11:35:32.25& -63:14:43.2&   -12.31&    1.8&    5.3&    6.8&    3.6&    3.2&       &       &       &  C \\
294.990-1.719& 11:39:22.88& -63:28:26.4&   -12.49&    3.8&   11.6&   12.7&    8.6&       &       &       &       &  F \\
298.262+0.739& 12:11:47.65& -61:46:20.9&   -30.39&    2.6&   10.0&   11.6&    6.7&    4.2&       &       &       &  C \\
299.772-0.005& 12:23:48.97& -62:42:25.3&    -6.87&    1.0&   12.4&   12.9&   11.9&    9.6&    9.3&    8.5&       &  A \\
305.200+0.019& 13:11:16.93& -62:45:55.1&   -32.15&    1.7&   31.4&   36.1&   29.7&   15.9&   25.8&       &       &  B \\
305.202+0.208& 13:11:10.49& -62:34:38.8&   -44.27&    2.2&   42.2&   83.7&   31.0&   14.3&  <15.7&       &       &  B \\
308.754+0.549& 13:40:57.60& -61:45:43.4&   -45.50&    1.5&    8.6&    9.5&    8.3&    4.9&    5.2&       &       &  B \\
308.918+0.123& 13:43:01.85& -62:08:52.2&   -54.46&    6.4&   24.0&   28.9&    3.5&       &       &       &       &  F \\
309.901+0.231& 13:51:01.05& -61:49:56.0&   -54.45&    0.6&   11.9&   15.8&   11.3&   11.0&   10.8&       &       &  A \\
309.921+0.479& 13:50:41.78& -61:35:10.2&   -58.94&    0.9&  145.5&  200.8&  138.9&  116.4&  108.3&       &       &  A \\
310.144+0.760& 13:51:58.43& -61:15:41.3&   -55.60&    1.5&   30.1&   30.8&   29.4&   19.6&   17.1&       &       &  A \\
311.643-0.380& 14:06:38.77& -61:58:23.1&   +32.38&   49.6&   10.7&    7.1&       &       &       &       &       &  F \\
312.071+0.082& 14:08:58.20& -61:24:23.8&   -34.35&    5.2&   34.3&   42.3&   11.5&       &       &       &       &  F \\
312.108+0.262& 14:08:49.31& -61:13:25.1&   -50.15&    3.3&   14.5&   16.7&    9.0&    2.5&       &       &       &  F \\
312.598+0.045& 14:13:15.03& -61:16:53.6&   -67.98&    2.1&   11.6&   15.1&    8.5&    7.0&       &       &       &  B \\
313.469+0.190& 14:19:40.94& -60:51:47.3&    -9.58&    1.8&   10.6&    5.6&   11.5&    5.8&       &       &       &  B \\
313.577+0.325& 14:20:08.58& -60:42:00.8&   -48.04&    1.6&   46.2&   78.9&   38.0&   32.8&   19.6&       &       &  A \\
313.767-0.863& 14:25:01.73& -61:44:58.1&   -43.04&    4.1&    8.2&    8.0&    5.5&       &       &       &       &  F \\
313.994-0.084& 14:24:30.78& -60:56:28.3&    -4.93&   53.4&   10.0&   10.0&       &       &       &       &       &  F \\
314.320+0.112& 14:26:26.20& -60:38:31.3&   -43.57&    1.2&   20.0&   28.7&   19.0&   13.6&   14.1&       &       &  A \\
316.359-0.362& 14:43:11.20& -60:17:13.3&    +3.15&    1.0&   76.8&   77.1&   73.8&   62.9&   50.1&       &       &  A \\
316.381-0.379& 14:43:24.21& -60:17:37.4&    -0.75&    1.2&    2.6&    5.2&    2.4&    2.0&    1.7&       &       &  B \\
316.412-0.308& 14:43:23.34& -60:13:00.9&    -5.88&    1.5&    5.7&    8.3&    4.5&    5.0&       &       &       &  B \\
316.640-0.087& 14:44:18.45& -59:55:11.5&   -20.64&    4.9&   40.2&   51.7&    2.9&    2.9&       &       &       &  F \\
316.811-0.057& 14:45:26.43& -59:49:16.3&   -45.76&    2.1&   40.0&   44.2&   31.4&   20.2&    4.3&    4.5&       &  A \\
317.466-0.402& 14:51:19.69& -59:50:50.7&   -39.00&    1.1&   20.7&   20.7&   19.8&   15.4&   13.6&   11.1&       &  A \\
317.701+0.110& 14:51:11.69& -59:17:02.1&   -42.34&    1.3&   18.5&   17.0&   18.5&   12.6&    6.6&   13.1&       &  B \\
318.050+0.087& 14:53:42.67& -59:08:52.4&   -51.64&    1.1&    8.4&    7.3&    9.3&    5.6&    5.1&    6.8&       &  B \\
318.948-0.196& 15:00:55.40& -58:58:52.1&   -34.87&    1.5&   94.0&  280.1&   86.7&   51.9&   54.4&   32.3&       &  A \\
320.231-0.284& 15:09:51.94& -58:25:38.5&   -62.45&    1.2&   19.1&   18.0&   20.0&   15.6&   11.4&    6.9&       &  B \\
320.780+0.248& 15:11:23.48& -57:41:25.1&    -5.29&    2.3&   12.7&   25.0&    5.2&    1.7&    4.5&       &       &  F \\
321.033-0.483& 15:15:52.63& -58:11:07.7&   -57.44&    1.1&   41.3&   43.1&   42.1&   30.9&   12.2&   33.9&       &  A \\
322.158+0.636& 15:18:34.64& -56:38:25.3&   -63.06&    1.6&  137.7&  241.4&  114.2&   98.9&  100.6&   33.7&       &  A \\
322.705-0.331& 15:25:47.52& -57:09:15.5&   -21.80&    1.2&    1.9&    2.0&    1.9&    1.3&       &       &       &  C \\
323.459-0.079& 15:29:19.33& -56:31:22.8&   -68.43&    1.8&   94.0&  232.9&   78.9&   54.0&   21.6&   32.9&       &  A \\
323.740-0.263& 15:31:45.45& -56:30:50.1&   -50.50&    1.9& 1215.3& 1982.9& 1030.3&  671.0&  302.6&       &       &  A \\
324.716+0.342& 15:34:57.47& -55:27:23.6&   -45.94&   46.9&  835.1&    6.7&       &       &       &       &       &  F \\
324.915+0.158& 15:36:51.17& -55:29:22.9&    -5.11&    1.7&    4.9&    7.0&    4.4&    2.3&    2.6&       &       &  C \\
326.475+0.703& 15:43:16.64& -54:07:14.6&   -38.56&    0.9&   30.4&   43.8&   29.0&   24.9&   27.0&       &       &  A \\
326.641+0.611& 15:44:33.33& -54:05:31.5&   -42.78&    1.5&   13.3&   24.6&   11.9&    8.3&    9.2&       &       &  B \\
326.662+0.520& 15:45:02.95& -54:09:03.1&   -40.84&    3.0&    5.9&    7.1&    3.4&    2.7&       &       &       &  F \\
326.859-0.677& 15:51:14.19& -54:58:04.8&   -58.14&    7.1&   12.7&   11.4&    4.5&       &       &       &       &  F \\
327.120+0.511& 15:47:32.73& -53:52:38.4&   -87.11&    2.9&   21.4&   27.9&   13.7&    4.8&       &       &       &  D \\
327.392+0.199& 15:50:18.48& -53:57:06.3&   -88.96&    0.0&    2.4&    2.5&    3.0&    4.4&    1.5&       &       &  C \\
327.402+0.445& 15:49:19.50& -53:45:13.9&   -82.99&    0.4&   45.3&   45.3&   46.2&   43.3&   46.3&       &       &  A \\
327.566-0.850& 15:55:47.61& -54:39:11.4&   -29.60&    1.5&    7.6&    9.1&    5.4&    4.9&       &       &       &  C \\
328.237-0.547& 15:57:58.28& -53:59:22.7&   -44.80&    1.7&  325.9&  434.0&  297.4&  215.8&  104.4&       &       &  A \\
328.254-0.532& 15:57:59.75& -53:58:00.4&   -37.06&    1.2&   18.2&   27.0&   18.7&   17.0&   14.5&       &       &  B \\
328.808+0.633& 15:55:48.45& -52:43:06.6&   -44.53&    1.3&   78.0&  140.4&   72.0&   78.0&       &       &   41.9&  A \\
329.029-0.205& 16:00:31.80& -53:12:49.6&   -36.19&    1.0&   39.7&   55.2&   41.8&   32.2&   17.1&       &   37.1&  A \\
329.066-0.308& 16:01:09.93& -53:16:02.6&   -47.43&    0.0&    5.2&    4.8&    5.7&    6.1&    8.0&       &       &  C \\
329.183-0.314& 16:01:47.01& -53:11:43.3&   -55.77&    0.0&    2.8&    2.4&       &    2.9&    3.5&       &       &  C \\
329.339+0.148& 16:00:33.13& -52:44:39.8&  -106.52&    1.6&   10.0&   10.3&    9.5&    6.4&       &       &       &  B \\
329.407-0.459& 16:03:32.65& -53:09:26.9&   -66.75&    1.2&   25.4&   32.9&   26.7&   19.5&    9.4&       &   19.4&  A \\
329.469+0.503& 15:59:40.71& -52:23:27.3&   -72.89&    0.9&    7.7&    6.4&    9.5&    6.9&       &       &    4.0&  B \\
329.610+0.114& 16:02:03.14& -52:35:33.5&   -60.25&    0.5&    8.5&    5.1&   11.2&    7.8&   24.6&       &    7.6&  B \\
329.719+1.164& 15:58:07.09& -51:43:32.6&   -75.70&    1.3&    3.8&    3.5&    4.3&    2.4&       &       &    2.2&  B \\
331.132-0.244& 16:10:59.77& -51:50:22.4&   -84.49&    3.6&    3.8&    4.2&    2.0&       &       &       &       &  F \\
331.278-0.188& 16:11:26.59& -51:41:56.7&   -77.63&    2.7&   25.6&   31.7&   17.8&    9.2&       &       &       &  D \\
331.342-0.346& 16:12:26.45& -51:46:16.4&   -67.10&    1.2&   27.5&   30.8&   26.1&   21.5&   23.6&       &   15.4&  A \\
331.425+0.264& 16:10:09.36& -51:16:04.5&   -88.78&    1.9&    7.5&    6.0&    7.6&    3.6&       &       &       &  C \\
331.442-0.187& 16:12:12.49& -51:35:10.1&   -88.52&    3.1&   18.5&   27.0&    5.0&    4.1&       &       &       &  F \\
331.542-0.066& 16:12:09.02& -51:25:47.6&   -84.57&   20.3&    3.8&    4.2&       &       &       &       &       &  F \\
331.556-0.121& 16:12:27.21& -51:27:38.2&   -97.21&    2.3&   18.1&   22.5&    9.8&    6.6&       &       &       &  F \\
331.710+0.603& 16:10:01.77& -50:49:32.3&   -75.00&    2.1&    8.1&    8.8&    7.0&    3.1&       &       &       &  C \\
332.094-0.421& 16:16:16.45& -51:18:25.7&   -58.58&    1.1&    6.5&    6.4&    6.9&    5.4&       &       &    3.1&  B \\
332.295+2.280& 16:05:41.72& -49:11:30.3&   -23.74&    0.8&   19.0&   20.0&   20.3&   18.3&       &       &   14.0&  A \\
332.813-0.701& 16:20:48.12& -51:00:15.6&   -58.23&    1.6&    3.5&    3.9&    3.3&    2.0&       &       &       &  D \\
332.963-0.679& 16:21:22.92& -50:52:58.5&   -46.03&    4.0&    4.5&    3.7&    6.9&       &       &       &       &  F \\
333.121-0.434& 16:20:59.71& -50:35:52.1&   -49.80&    1.4&    8.4&    8.6&    8.0&    7.1&       &       &       &  C \\
333.128-0.560& 16:21:35.38& -50:40:56.5&   -56.83&    1.9&    4.9&    4.6&    5.0&    3.6&       &       &       &  D \\
333.163-0.101& 16:19:42.67& -50:19:53.2&   -95.19&    2.6&    6.7&    7.0&    6.2&    1.6&       &       &       &  D \\
333.184-0.091& 16:19:45.62& -50:18:35.0&   -81.76&    2.2&    3.8&    4.2&    2.4&       &       &       &       &  D \\
333.315+0.105& 16:19:29.01& -50:04:41.3&   -43.92&    1.2&    9.4&    8.4&   10.5&    6.7&       &       &    4.7&  B \\
333.466-0.164& 16:21:20.18& -50:09:48.6&   -42.08&    0.9&   14.3&   15.0&   14.9&   12.9&   15.2&       &    8.1&  B \\
333.562-0.025& 16:21:08.80& -49:59:48.0&   -35.41&    1.4&   17.7&   20.7&   17.2&   12.3&   12.1&       &    4.5&  B \\
333.646+0.058& 16:21:09.14& -49:52:45.9&   -87.56&    1.4&    3.9&    4.4&    3.6&    3.2&       &       &       &  C \\
333.683-0.437& 16:23:29.78& -50:12:08.6&    -5.64&   41.7&    5.7&    6.3&       &       &       &       &       &  F \\
333.931-0.135& 16:23:14.83& -49:48:48.9&   -36.98&    2.1&    7.4&    8.2&    6.2&    3.7&       &       &       &  D \\
334.635-0.015& 16:25:45.73& -49:13:37.4&   -30.13&    2.8&   14.4&   13.9&   10.2&    5.2&       &       &       &  D \\
335.060-0.427& 16:29:23.13& -49:12:27.1&   -47.09&    3.4&   16.7&   21.9&    8.6&    3.7&       &       &       &  D \\
335.426-0.240& 16:30:05.58& -48:48:44.8&   -50.68&    2.5&   22.5&   30.6&   10.9&   14.5&       &       &       &  D \\
335.556-0.307& 16:30:55.96& -48:45:50.0&  -116.44&    3.9&    9.1&    9.9&    2.6&       &       &       &       &  F \\
335.585-0.285& 16:30:57.28& -48:43:39.7&   -44.02&    2.0&    4.7&    5.6&    3.9&    2.7&    2.5&       &       &  C \\
335.726+0.191& 16:29:27.37& -48:17:53.2&   -44.45&    3.4&    5.5&    7.8&    3.8&       &       &       &       &  D \\
335.789+0.174& 16:29:47.33& -48:15:51.7&   -47.69&    4.6&   46.0&   56.8&   14.9&    5.5&       &       &       &  D \\
336.018-0.827& 16:35:09.30& -48:46:46.8&   -41.37&    5.1&   30.2&   39.5&    6.7&   <1.9&       &       &       &  F \\
336.358-0.137& 16:33:29.17& -48:03:43.9&   -73.51&    2.1&    8.6&    9.6&    3.3&    2.7&       &       &       &  C \\
336.433-0.262& 16:34:20.22& -48:05:32.2&   -92.74&    4.5&   16.6&   18.2&    5.8&       &       &       &       &  F \\
336.496-0.271& 16:34:38.02& -48:03:03.9&   -24.17&   22.4&    6.4&    6.0&       &       &       &       &       &  F \\
336.822+0.028& 16:34:38.28& -47:36:32.2&   -76.85&    4.9&   13.5&   14.1&    5.2&       &       &       &       &  F \\
336.830-0.375& 16:36:26.19& -47:52:31.1&   -24.52&   23.8&    6.2&    7.1&       &       &       &       &       &  F \\
336.864+0.005& 16:34:54.44& -47:35:37.3&   -75.88&    5.7&   18.3&   18.6&    3.8&       &       &       &       &  F \\
336.941-0.156& 16:35:55.19& -47:38:45.4&   -64.91&    5.4&   11.7&   13.4&    3.6&       &       &       &       &  F \\
336.983-0.183& 16:36:12.41& -47:37:58.2&   -80.61&   25.4&    7.6&    9.3&       &       &       &       &       &  F \\
336.994-0.027& 16:35:33.98& -47:31:11.7&  -125.84&    4.0&   12.8&   15.3&    5.2&       &       &       &       &  F \\
337.052-0.226& 16:36:40.12& -47:36:38.2&   -77.29&    2.4&    5.7&    6.8&    3.9&    1.6&       &       &       &  D \\
337.153-0.395& 16:37:48.86& -47:38:56.5&   -49.45&    1.8&    7.6&    9.3&    6.7&    4.1&    4.2&       &       &  C \\
337.201+0.114& 16:35:46.56& -47:16:16.7&   -54.55&    1.9&    3.5&    3.7&    2.9&       &       &       &       &  D \\
337.388-0.210& 16:37:56.01& -47:21:01.2&   -56.04&    2.4&    6.6&    8.1&    4.2&    2.3&       &       &       &  D \\
337.404-0.402& 16:38:50.51& -47:28:00.3&   -39.88&    1.8&    9.2&   10.9&    8.8&    4.5&    3.9&       &       &  C \\
337.613-0.060& 16:38:09.54& -47:04:59.9&   -43.00&    2.7&    8.7&   11.4&    4.8&    4.8&       &       &       &  D \\
337.632-0.079& 16:38:19.13& -47:04:53.5&   -56.90&    3.6&    4.6&    5.0&    2.8&       &       &       &       &  F \\
337.705-0.053& 16:38:29.63& -47:00:35.5&   -54.72&    4.0&   44.3&   64.3&   14.4&    6.2&       &       &       &  D \\
337.920-0.456& 16:41:06.05& -47:07:02.1&   -38.74&    0.0&   10.0&   12.4&    9.7&   10.9&   14.1&       &   10.5&  B \\
338.075+0.012& 16:39:39.04& -46:41:28.0&   -52.96&    2.9&    5.9&    6.3&    4.3&   <0.8&       &       &       &  D \\
338.287+0.120& 16:40:00.13& -46:27:37.1&   -40.06&    1.4&   11.4&   11.7&   11.5&    7.8&       &       &       &  C \\
338.432+0.058& 16:40:49.79& -46:23:37.0&   -30.13&    3.5&   12.7&   15.8&    6.2&    1.8&       &       &       &  F \\
338.461-0.245& 16:42:15.50& -46:34:18.4&   -52.00&    2.6&   11.3&   13.9&    9.2&    2.8&       &       &       &  D \\
338.472+0.289& 16:39:58.91& -46:12:35.4&   -29.87&    1.2&    0.5&    0.4&    0.5&    0.3&       &       &       &  N \\
338.497+0.207& 16:40:25.89& -46:14:43.5&   -28.21&    1.3&    1.5&    1.5&    1.4&    1.1&       &       &       &  D \\
338.561+0.218& 16:40:37.96& -46:11:25.8&   -39.09&    2.2&   16.5&   16.2&   13.3&    5.6&       &       &       &  C \\
338.566+0.110& 16:41:07.03& -46:15:28.3&   -74.91&    3.1&    5.4&    5.6&    3.6&       &       &       &       &  F \\
338.850+0.409& 16:40:54.29& -45:50:52.0&   -57.61&    0.0&    0.7&    0.7&    0.7&    0.9&       &       &       &  N \\
338.875-0.084& 16:43:08.25& -46:09:12.8&   -41.37&    3.0&    5.5&    6.3&    3.0&       &       &       &       &  F \\
338.902+0.394& 16:41:10.06& -45:49:05.4&   -30.66&    0.0&    1.1&    0.9&    1.4&       &       &       &       &  N \\
338.920+0.550& 16:40:34.01& -45:42:07.1&   -61.30&    1.5&   30.5&   38.2&   25.4&   21.1&       &    9.8&    7.7&  B \\
338.925+0.634& 16:40:13.56& -45:38:33.2&   -60.77&    1.0&   17.5&   19.5&   18.1&   13.8&       &   11.8&    8.7&  B \\
338.935-0.062& 16:43:16.01& -46:05:40.2&   -41.90&    1.7&   15.0&   16.7&   13.7&    7.9&       &    5.0&       &  B \\
339.053-0.315& 16:44:48.99& -46:10:13.0&  -111.88&    4.0&   26.4&   46.7&    4.4&    5.9&       &       &       &  D \\
339.622-0.121& 16:46:05.99& -45:36:43.3&   -33.20&    0.0&   21.7&   16.9&   24.7&   26.9&       &   23.2&   21.6&  A \\
339.681-1.208& 16:51:06.21& -46:16:02.8&   -21.44&    2.3&   14.1&   21.6&   11.2&    4.6&       &       &       &  C \\
339.762+0.054& 16:45:51.56& -45:23:32.6&   -51.03&   18.0&    3.0&    3.3&       &       &       &       &       &  F \\
339.884-1.259& 16:52:04.67& -46:08:34.1&   -35.58&    0.0&  341.9&  299.6&  408.2&  406.0&       &  373.1&  521.1&  A \\
339.949-0.539& 16:49:07.98& -45:37:58.3&   -91.42&    2.8&   20.0&   24.2&   13.2&    4.7&       &       &       &  D \\
339.986-0.425& 16:48:46.31& -45:31:51.3&   -88.43&    2.9&   16.0&   26.6&    7.9&    5.3&       &       &       &  C \\
340.054-0.244& 16:48:13.89& -45:21:43.3&   -60.95&    0.4&   13.8&   12.1&   17.3&   14.8&       &   10.4&   14.5&  B \\
340.249-0.046& 16:48:05.18& -45:05:08.4&  -126.37&    2.8&    3.5&    3.3&    7.4&       &       &       &       &  D \\
340.785-0.096& 16:50:14.84& -44:42:26.3&  -108.02&    4.2&  157.8&  246.0&   50.3&   13.1&       &       &       &  F \\
340.970-1.022& 16:54:57.32& -45:09:05.2&   -32.41&    4.1&    5.4&    5.6&    5.1&       &       &       &       &  F \\
341.218-0.212& 16:52:17.84& -44:26:52.1&   -44.44&    3.6&   64.0&   91.5&   27.6&    3.7&       &       &       &  F \\
342.484+0.183& 16:55:02.30& -43:12:59.8&   -42.24&    3.8&   16.8&   24.4&    5.3&    7.9&       &       &       &  C \\
343.354-0.067& 16:59:04.24& -42:41:34.6&  -117.76&    3.7&    6.4&    6.1&   14.3&       &       &       &       &  F \\
344.227-0.569& 17:04:07.78& -42:18:39.5&   -23.98&    1.7&   42.3&   40.3&   41.6&   23.1&       &       &       &  B \\
344.421+0.045& 17:02:08.77& -41:46:58.5&   -71.49&    3.8&    8.5&    9.9&    3.7&       &       &       &       &  F \\
345.003-0.223& 17:05:10.89& -41:29:06.2&   -22.93&    2.7&   38.8&   58.2&   23.7&   12.2&       &       &       &  C \\
345.010+1.792& 16:56:47.58& -40:14:25.8&   -21.71&    1.0&  105.1&   92.2&  125.3&  101.9&       &   37.7&  107.8&  A \\
345.505+0.348& 17:04:22.91& -40:44:21.7&   -19.07&    0.3&   82.8&   84.2&   92.8&  100.6&       &   72.9&    7.9&  A \\
346.036+0.048& 17:07:20.02& -40:29:49.0&    -6.52&    3.8&    3.7&    3.9&    1.4&       &       &       &       &  F \\
346.480+0.221& 17:08:00.11& -40:02:15.9&   -18.89&    1.7&    9.5&   11.2&    8.7&    5.3&       &       &       &  C \\
347.631+0.211& 17:11:36.05& -39:07:07.0&   -91.94&    2.2&    1.8&    1.8&    1.9&       &       &       &       &  D \\
348.550-0.979& 17:19:20.41& -39:03:51.6&   -10.46&    0.6&   11.7&   12.0&   14.2&    9.9&       &   10.7&       &  B \\
348.617-1.162& 17:20:18.65& -39:06:50.8&   -11.52&    1.1&   21.9&   20.8&   22.9&   18.9&       &   10.0&       &  B \\
348.703-1.043& 17:20:04.06& -38:58:30.9&    -7.39&    0.0&   12.6&   10.7&   15.8&   12.2&       &       &       &  B \\
348.884+0.096& 17:15:50.13& -38:10:12.4&   -74.38&    1.2&    4.6&    4.9&    5.2&       &       &       &       &  D \\
349.092+0.105& 17:16:24.74& -37:59:47.2&   -76.58&    2.8&    6.4&    7.5&    4.7&    4.1&       &       &       &  D \\
350.105+0.083& 17:19:27.01& -37:10:53.3&   -74.21&    4.1&    5.6&    6.3&    2.7&       &       &       &       &  F \\
350.299+0.122& 17:19:50.87& -36:59:59.9&   -61.99&    4.1&   10.2&   12.0&    5.9&       &       &       &       &  F \\
350.686-0.491& 17:23:28.63& -37:01:48.8&   -13.89&    0.5&    2.7&    2.1&    3.3&    3.1&       &    1.8&       &  B \\
351.161+0.697& 17:19:57.50& -35:57:52.8&    -5.29&   17.8&    3.2&    3.3&       &       &       &       &       &  F \\
351.382-0.181& 17:24:09.58& -36:16:49.3&   -59.63&   17.2&    9.7&    6.9&       &       &       &       &       &  F \\
351.417+0.645& 17:20:53.37& -35:47:01.2&   -10.38&    1.6&  243.0&  361.7&  299.4&  115.9&       &  120.7&       &  A \\
351.581-0.353& 17:25:25.12& -36:12:46.1&   -94.23&    1.5&    7.6&    9.4&    7.2&    4.7&       &       &       &  D \\
351.688+0.171& 17:23:34.52& -35:49:46.3&   -36.19&    1.1&   13.8&   10.9&   16.6&   11.4&       &    7.0&       &  B \\
351.775-0.536& 17:26:42.57& -36:09:17.6&    +1.56&    4.2&   37.9&   37.1&   21.5&       &       &       &       &  F \\
352.133-0.944& 17:29:22.23& -36:05:00.2&   -16.00&    3.4&    5.1&    5.2&    4.5&       &       &       &       &  F \\
352.630-1.067& 17:31:13.91& -35:44:08.7&    -3.26&    1.5&   68.9&   60.7&   73.8&   43.7&       &   19.6&       &  A \\
353.273+0.641& 17:26:01.58& -34:15:15.4&    -4.50&    2.5&    2.1&    2.1&    1.6&       &       &       &       &  D \\
353.410-0.360& 17:30:26.18& -34:41:45.6&   -17.33&    0.0&   62.2&   43.9&   74.7&  104.5&       &       &       &  N \\
353.429-0.090& 17:29:23.48& -34:31:50.3&   -61.48&   19.8&    3.9&    4.4&       &       &       &       &       &  F \\
353.464+0.562& 17:26:51.53& -34:08:25.7&   -50.42&    2.4&    3.3&    3.5&    2.3&       &       &       &       &  D \\
354.615+0.472& 17:30:17.13& -33:13:55.1&   -24.25&    3.3&   51.2&   52.4&   28.9&   12.6&       &       &       &  D \\
354.724+0.300& 17:31:15.55& -33:14:05.7&   +92.61&    0.0&    3.7&    4.0&    4.0&    6.0&       &       &       &  C \\
355.346+0.149& 17:33:28.91& -32:47:49.5&   +19.92&   15.5&    2.3&    2.5&       &       &       &       &       &  F \\
356.662-0.263& 17:38:29.16& -31:54:38.8&   -53.84&   19.4&    3.8&    4.1&       &       &       &       &       &  F \\
357.967-0.163& 17:41:20.26& -30:45:06.9&    -3.09&    4.1&    9.4&    9.3&    8.0&       &       &       &       &  F \\
358.263-2.061& 17:49:37.63& -31:29:18.0&    +5.90&    3.3&    6.1&    6.5&    4.9&       &       &       &       &  F \\
358.371-0.468& 17:43:31.95& -30:34:11.0&    +1.31&    1.0&   12.6&   14.8&   11.5&   13.5&       &       &       &  C \\
358.386-0.483& 17:43:37.83& -30:33:51.5&    -5.98&    0.4&    3.1&    2.8&    3.6&    3.0&       &    2.4&       &  B \\
358.460-0.391& 17:43:26.76& -30:27:11.3&    +1.20&    0.6&   12.7&   16.5&   13.5&   12.8&       &   10.4&       &  B \\
358.809-0.085& 17:43:05.40& -29:59:45.8&   -56.11&    0.0&    2.3&    2.5&    2.7&    1.3&       &    4.3&       &  C \\
358.841-0.737& 17:45:44.29& -30:18:33.6&   -20.73&    0.8&    5.7&    5.9&    6.2&    4.5&       &       &       &  C \\
358.931-0.030& 17:43:10.02& -29:51:45.8&   -15.90&    1.2&    2.0&    2.1&    2.1&    2.1&       &       &       &  C \\
359.138+0.031& 17:43:25.69& -29:39:17.4&    -3.79&    8.7&    2.6&    2.5&       &       &       &       &       &  F \\
359.436-0.104& 17:44:40.60& -29:28:16.0&   -46.70&    2.7&   18.0&   21.8&    2.5&   10.3&       &       &       &  D \\
359.615-0.243& 17:45:39.09& -29:23:30.0&   +19.56&    1.7&   17.5&   14.9&   17.5&   12.3&       &       &       &  B \\
\end{longtable}

\captionsetup{type=figure} % optional, makes it a figure for list of figures
\captionof{figure}{
  Compactness plot for each maser observed in this survey. The x-axis is $uv$-distance with units M$\lambda$, shown with ${uv}^2$ scaling. The y-axis is flux density in Jy, with $\log_{10}$ scaling. Blue circles indicate detected flux density, while red points indicate when the maser channel was observed at a given baseline length, but was not detected. The black-dashed line shows the detection threshold. The blue solid line shows the least squares fit used to determine the size of the emission region. Only the most ``most-compact'' 2~kHz velocity channel is shown per maser, with the velocity given in the legend. All other velocity channels are equally or less compact.}

\newcount\figcount
\newcount\cols
\cols=4 % number of columns
\figcount=0
\newdimen\imgwidth
\imgwidth=0.23\textwidth

\newread\figfile
\openin\figfile=filelist.txt

\begingroup
\noindent
\loop
  \read\figfile to \filename
  \unless\ifeof\figfile
    \ifx\filename\empty
      % skip empty line
    \else
      \includegraphics[width=\imgwidth]{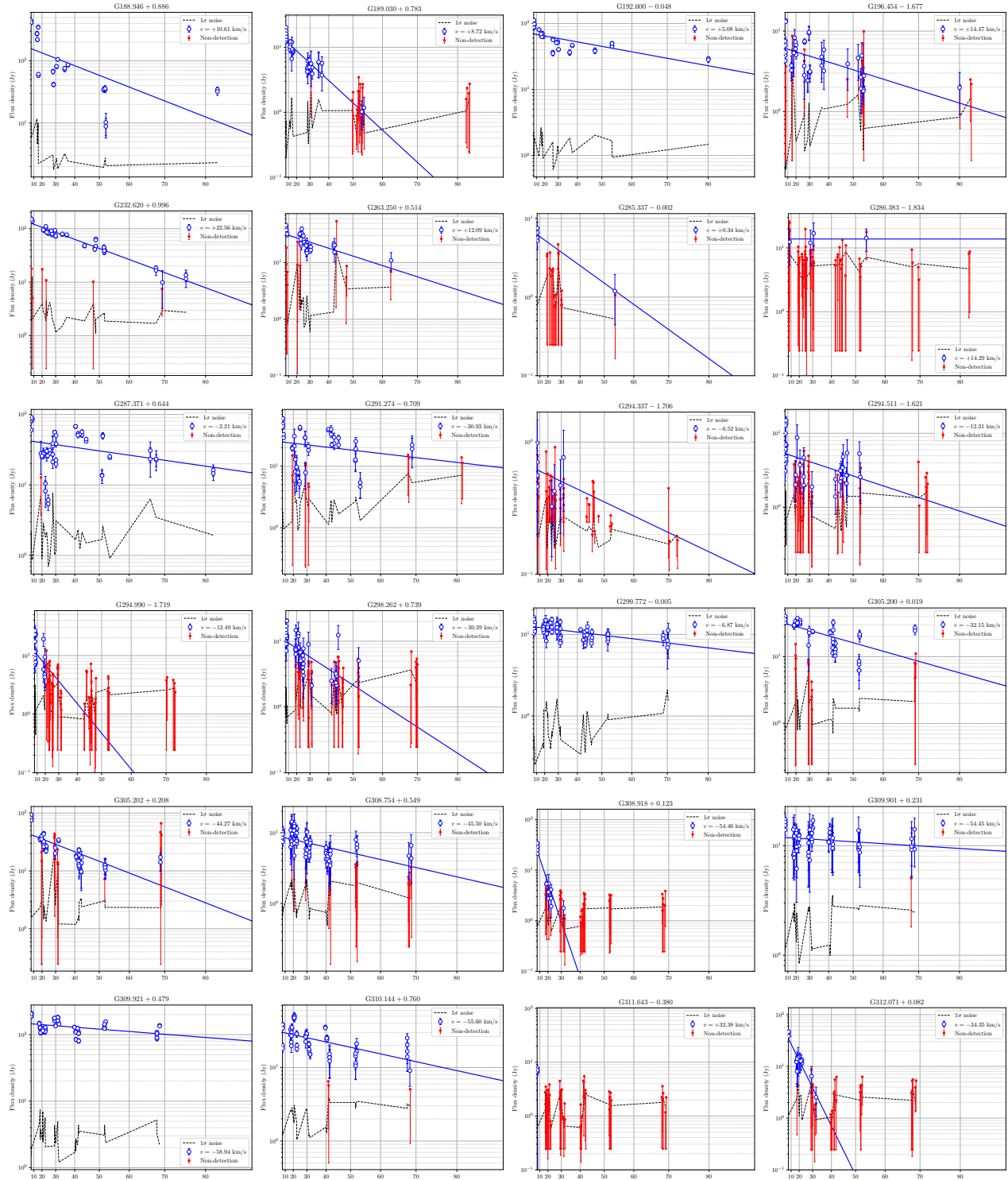}%
      \advance\figcount by 1
      \ifnum\figcount=\cols
        \\[2pt]
        \figcount=0
      \fi
    \fi
  \repeat
\endgroup

\closein\figfile

\end{document}